\documentclass[10pt]{iopart}
\pdfoutput=1
\usepackage{graphicx}

\usepackage{iopams}
\usepackage{comment}
\usepackage{xcolor}

\begin{document}


\title{Bifurcation and Criticality}
\author{Indrani Bose}
\address{Department of Physics, Bose Institute, Kolkata-700009, India}
\ead{indrani@jcbose.ac.in}
\author{Sayantari Ghosh}
\address{Department of Physics, National Institute of Technology, Durgapur-713209, India}
\ead{sayantari@gmail.com}
\vspace{10pt}
\begin{indented}
\item[]December 2018
\end{indented}

\begin{abstract}
Equilibrium and nonequilibrium systems exhibit power-law singularities close to their critical and bifurcation points respectively. A recent study has shown that biochemical nonequilibrium models with positive feedback belong to the universality class of the mean-field Ising model. Through a mapping between the two systems, effective thermodynamic quantities like temperature, magnetic field and order parameter can be expressed in terms of biochemical parameters. In this paper, we demonstrate the equivalence using a simple deterministic approach. As an illustration we consider a model of population dynamics exhibiting the Allee effect for which we determine the exact phase diagram. We further consider a two-variable model of positive feedback, the genetic toggle, and discuss the conditions under which the model belongs to the mean-field Ising universality class. In the biochemical models, the supercritical pitchfork bifurcation point serves as the critical point. The dynamical behaviour predicted by the two models is in qualitative agreement with experimental observations and opens up the possibility of exploring critical point phenomena in laboratory populations and synthetic biological circuits.
\end{abstract}
\vspace{2pc}

%

\section{Introduction}
Bifurcations in dynamical systems have certain similarities with thermodynamic phase transitions in equilibrium systems. In a magnetic solid, the transition from a paramagnetic to a ferromagnetic phase occurs at a critical temperature with the two phases having distinct physical properties. For example, the spontaneous magnetization, i.e., the magnetization in the absence of an external magnetic field,  is zero in the paramagnetic phase, whereas it has a finite value in the ferromagnetic phase. The thermodynamic quantity, spontaneous magnetization, serves as the order parameter of the transition. A bifurcation in a dynamical system occurs at a critical parameter value and involves changes in the number of steady states and/or their stability properties. Steady states are, in general, attained under nonequilibrium conditions. In a steady state, the  temporal rate of change of a dynamical variable is zero. The change in dynamical regime from, say, monostability to bistability at a bifurcation point is akin to a phase transition \cite{erez2017criticality, sornette2006transitions, qian2016framework,munoz2018colloquium}. The major difference between a bifurcation and a thermodynamic phase transition is that the former occurs in a finite-dimensional state space with the dynamical system having a finite number of degrees of freedom whereas the latter requires the thermodynamic limit, i.e., the system size going to infinity for the emergence of critical point singularities in thermodynamic quantities. The asymptotic limit of time  $t\longrightarrow \infty$ is required for a dynamical system to attain the attractor state. This asymptotic limit in the case of bifurcations is analogous to the thermodynamic limit in the case of equilibrium phase transitions \cite{lesne2011scale}. Critical singularities appear at the bifurcation point only in this limit. The characteristic/return/relaxation  time $T_R$ of a dynamical system is the time required by the system to attain the attractor state after being weakly perturbed from it. The return time diverges as the bifurcation point is approached similar to the critical slowing down occurring close to the thermodynamic critical points \cite{erez2017criticality, lesne2011scale, sole2011phase}. Thermal fluctuations are an essential component of thermodynamic phase transitions.  Bifurcations, on the other hand, are associated with deterministic dynamics. Inclusion of noise (fluctuations) in the dynamics has the effect of smearing the transition between different dynamical regimes. Other analogies between bifurcations and phase transitions will be brought out in appropriate Sections of the paper.\\
\noindent Positive feedback in a biochemical system is well-known to promote bistability with the transition from a monostable to a bistable dynamical regime occurring at a bifurcation point \cite{pomerening2008uncovering,bose2012culture}. Recently, Erez et al. \cite{erez2017criticality}  demonstrated the equivalence between biochemical models with positive feedback and the Ising mean-field model. The mapping is strictly valid close to the bifurcation point and the critical point of the dynamical and Ising models respectively. Thermodynamic quantities exhibit scaling and power-law singularities in the proximity of the equilibrium critical point. The feedback models close to their bifurcation points exhibit similar scaling with the  critical behaviour belonging to the Ising mean-field class. The mapping between the equilibrium and nonequilibrium models makes it possible to define the effective order parameter, temperature, external field and heat capacity in terms of the biochemical parameters. The study by Erez et al. \cite{erez2017criticality}  utilises the stochastic birth-death model of a well-mixed system to explore the analogies between the critical behaviour of equilibrium Ising  and nonequilibrium models with positive feedback. The models considered by them are single-variable models. In this paper, we adopt a simpler approach of demonstrating the equivalence between the Ising and biochemical models. In Section 1, we outline the basic steps of the mapping assuming deterministic dynamics for the positive feedback models. In Section 2, a single-variable  model of population dynamics exhibiting the Allee effect is considered as an illustration of the mapping procedure. The phase diagram of the model is determined and its analogy with thermodynamic phase diagrams pointed out. The conditions for the validity of the mapping for a two-variable dynamical model, the genetic toggle, are discussed in Section 3. Section 4 contains some concluding remarks. 
\begin{figure}
\begin{center}
\begin{minipage}[c]{0.99\linewidth}
\includegraphics[width=13.2cm]{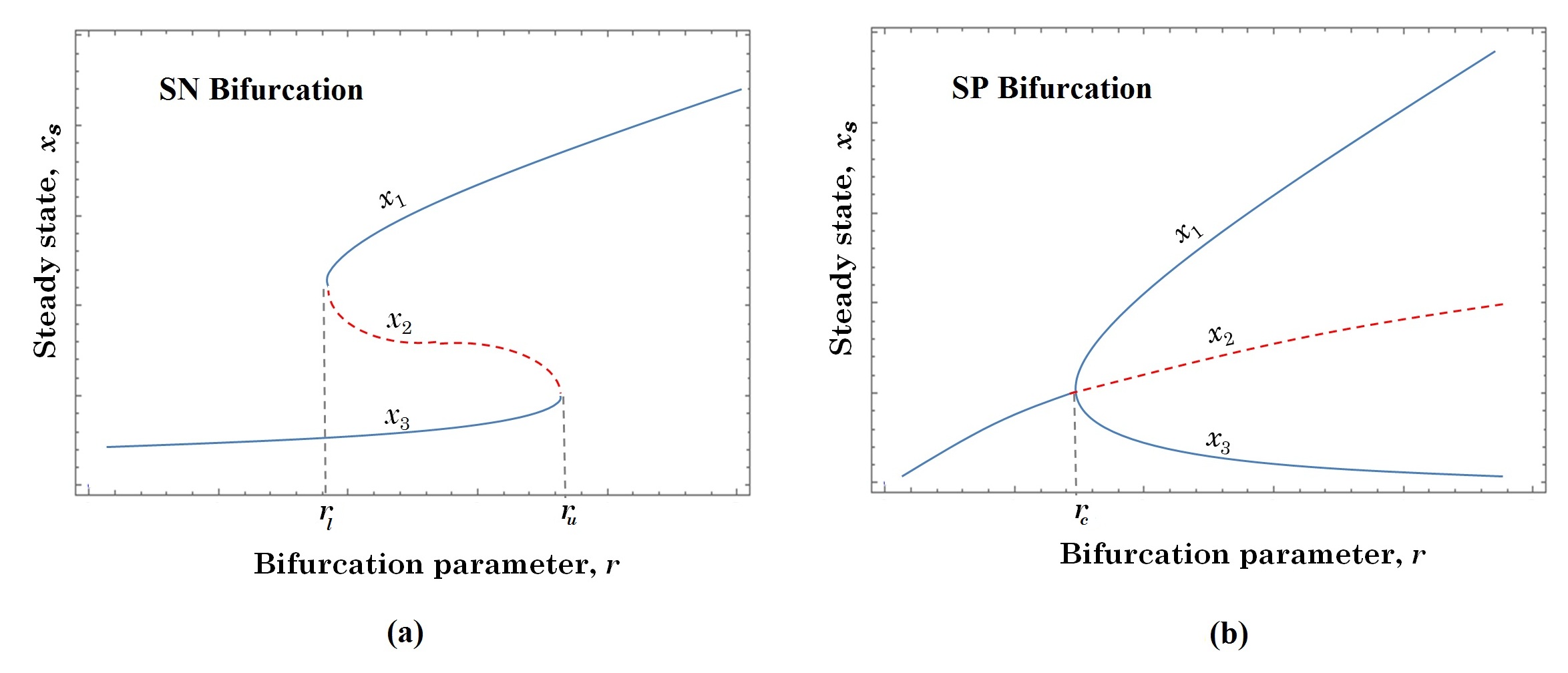}
\end{minipage}

\caption{Steady state values of $x_s$ versus the bifurcation parameter $r$ in the cases of (a) SN and (b) SP bifurcations.  In the region of bistability, the steady states are $x_1$ (stable), $x_2$ (unstable) and $x_3$ (stable). The solid (dashed) branches represent stable (unstable) steady states. The parameters $r_l$ and $r_u$ in (a) represent the lower and upper SN bifurcation points whereas the parameter $r=r_c$ in (b) represent the SP bifurcation point.}
\end{center}
\label{snsp}
\end{figure}
\section{Equivalence between mean-field criticality and bifurcation}
We consider a single-variable dynamical system the time evolution of which is governed by the differential rate equation

\begin{equation}
\frac{dx}{dt}=F(x,r)
\end{equation}
\noindent where   $x$  represents the dynamical variable and $r$  is the bifurcation parameter. In the steady state, $dx/dt=0$.  The bistable regime is reached via either a saddle-node (SN) or a supercritical pitchfork (SP) bifurcation \cite{strogatz2018nonlinear}. In models with positive feedback, a pair of SN bifurcations separates the bistable region from two regions of monostability, i.e., the dynamical system enters the bistable regime at the lower bifurcation  $r_l$ and exits the regime at the upper bifurcation point $r_u$. In the case of a SP transition, there is a single bifurcation point at which the transition between monostability and bistability occurs. Figures 1(a) and 1(b) show the plots of the steady state $x_s$ versus the bifurcation parameter $r$ in the cases of SN and SP bifurcations respectively.  In the region of bistability, the condition $F(x,r)=0$  is satisfied by three physical roots $x_1,x_2,x_3$ representing the steady states. The steady state $x_2$ is unstable and separates the states $x_1$ and $x_3$ which are stable. At a bifurcation point, $r=r_c$ ($r_c=r_l$ or $r_u$ for SN bifurcation),  a stable steady state $x^*$ loses stability. In the case of SN bifurcation, the steady state $x^*$  (could be either $x_1$ or $x_3$) merges with the unstable steady state $x_2$ at the bifurcation point and both the states cease to exist as physical states beyond the bifurcation point. Thus $x^*$ is a double root of $F(x,r)=0$ at $r=r_c$ $(r_c=r_l$ or $r_u)$. The conditions for SN bifurcation are \cite{karantonis2002easy}:
\begin{equation}
F(x^*,r_c )=0,\frac{\partial F(x^*,r_c )}{\partial x}=0,   \frac{\partial F(x^*,r_c )}{\partial r}\neq 0,\frac{\partial^2 F(x^*,r_c )}{\partial x^2}\neq 0
\label{eq_sn}
\end{equation}

\noindent The second condition in Equation (\ref{eq_sn}) can alternatively be written as $\lambda=0$ where $\lambda$ is the stability parameter. 
In the case of the SP bifurcation, $x^*$ is a triple root of $F(x,r)=0$ at $r=r_c$. The conditions for the SP bifurcation are \cite{karantonis2002easy}:
\begin{eqnarray}
F(x^*,r_c )&=&0,\frac{\partial F(x^*,r_c )}{\partial x}=0,   \frac{\partial F(x^*,r_c )}{\partial r} = 0,\\
\frac{\partial^2 F(x^*,r_c )}{\partial x^2}&=& 0, \frac{\partial^2 F(x^*,r_c )}{\partial x \partial r} \neq 0, \frac{\partial^3 F(x^*,r_c )}{\partial x^3}\neq 0 \nonumber
\label{eq_sp}
\end{eqnarray}
\noindent The differential rate equation of a dynamical system can be reduced to the normal form close to a bifurcation point \cite{karantonis2002easy,guckenheimer2013nonlinear}. This is achieved by expanding $F(x,r)$ around $x^*$ and $r_c$ to lowest contributing orders and introducing rescaled coordinates $X$ and $R$. The normal form of a SP  bifurcation is given by
\begin{equation}
\frac{dX}{dt}=RX-X^3
\label{sp_norm}
\end{equation}
\noindent The rate equation is reflection symmetric, i.e., has the same form under the transformation $X\longrightarrow-X$. The symmetry is lost in the case of imperfect SP bifurcation with the rate equation
\begin{equation}
\frac{dX}{dt}=RX-X^3+H
\label{imperfect}
\end{equation}
\noindent where $H$ is the imperfection parameter. The steady state equation
\begin{equation}
RX-X^3+H=0\label{imper}
\end{equation}
\noindent is identical in form to the mean-field (MF) equation of state of the Ising model close to the critical point \cite{barenblatt1992n},
\begin{equation}
h-\theta m-\frac{1}{3} m^3=0
\label{ising}
\end{equation}

\begin{figure}
\begin{minipage}[c]{0.99\linewidth}
\begin{center}
\includegraphics[width=13.2cm]{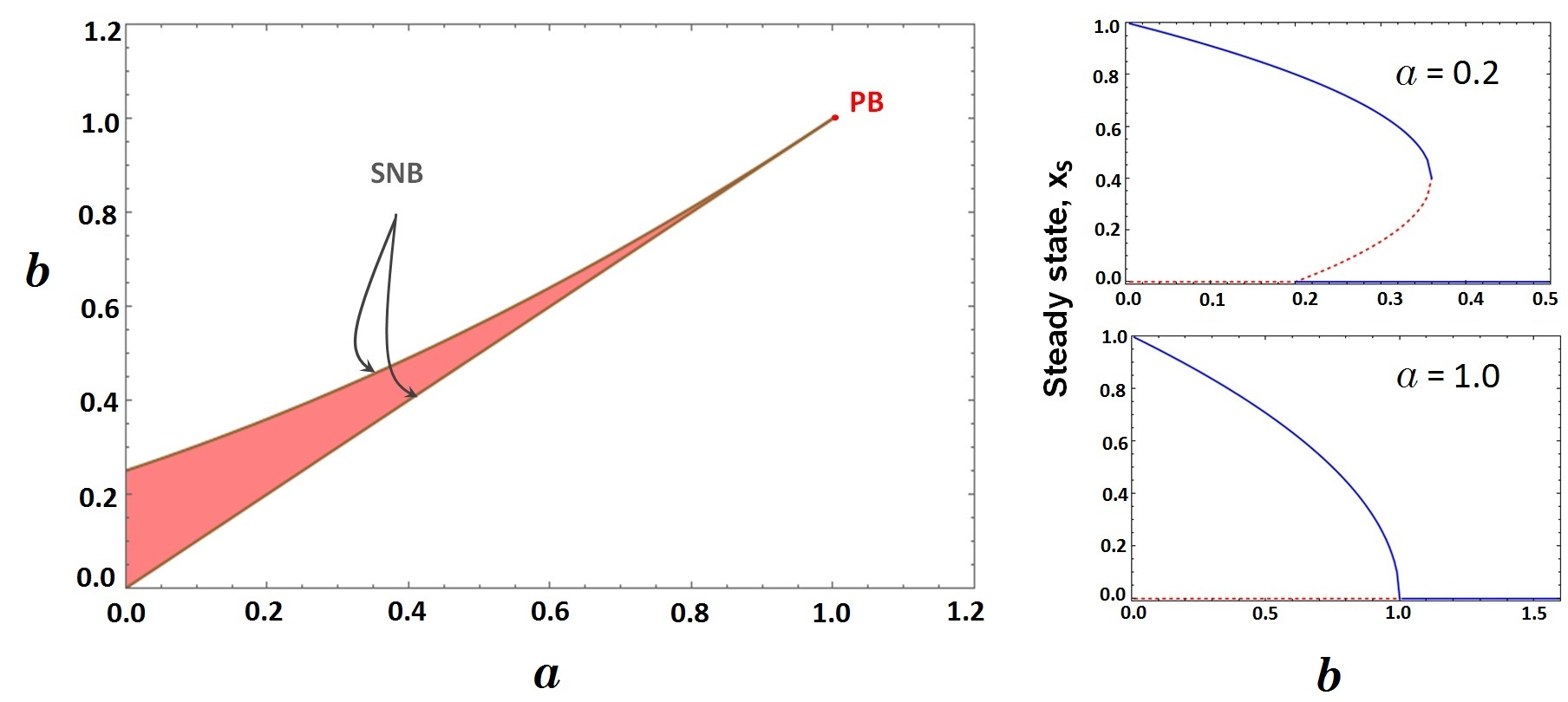}
\end{center}
\end{minipage}
\caption{(Left panel) The $b-a$ phase diagram of the Allee model. The shaded region is the region of bistability bounded by lines of SN bifurcations (SNBs) and terminating at the SP bifurcation point PB corresponding to $a = 1; b = 1 $. (Right panel) Steady state $x_s$ versus the parameter $b$ for $a = 0.2$ (upper plot) and $a = 1.0$ (lower plot). The solid (dashed) lines represent stable (unstable) steady states. The upper plot corresponds to SN bifurcation at which a discontinuous jump to population extinction takes place. In the lower plot, there is a continuous transition from a finite to zero population density at the SP bifurcation point.}
\label{alleephase}
\end{figure}
\noindent where $m$ is the average magnetization per spin, $\theta=(T-T_c )/T_c$   is the reduced temperature with $T_c$ being the critical temperature and $h=\frac{h_{ext}}{2dJ_{ex}}$ is the dimensionless magnetic field. The quantities $d,J_{ex},h_{ext}$ are the dimension of the system, the strength of the interaction between the Ising spins and the external magnetic field respectively. The critical point for the Ising model is  $(\theta=0,h=0)$ at which the order parameter $m=0$. The analogous SP bifurcation point is $(R=0,H=0)$ at  which the triple root $X^*=0$ describes the steady state. Given the functional form $F(x,r)$ (Equation(1)) of a dynamical system with positive feedback, the SP bifurcation point is identified if the conditions set in Equation (3) are satisfied. \\  
We  express the Ising parameters  in terms of the positive feedback model parameters by reducing the steady state equation $F(x^* )=0$ near the SP bifurcation point to the normal form   (Equation (\ref{imper})) of the imperfect SP bifurcation. Taylor expansion of $F(x^* )$ is carried out around the point $x_c$ to third order. The value of $x_c$ is fixed by the condition  $F^{''} (x_c )=0$ so that a second order term does not appear in the Taylor expansion consistent with  the normal forms  in   Equations (6) and (7). Thus, we get
\begin{equation}
F(x_c )+F^{'} (x_c )  (x^*-x_c )+F^{'''} (x_c ) \frac{(x^*-x_c )^3}{3!}=0
\end{equation}
\noindent This condition can be put in the normal form, Equation (\ref{imper}), with
\begin{equation}
X=\frac{(x^*-x_c )}{x_c} ,H=-\frac{6F(x_c )}{F^{'''} (x_c ) x_c^3 },R= -\frac{6F^{'}(x_c )}{F^{'''} (x_c ) x_c^2 }
\label{mag}
\end{equation}
\noindent Comparing Equations (\ref{imper}) and (\ref{ising}), we obtain 
\begin{equation}
m=\frac{(x^*-x_c )}{x_c} ,h=-\frac{2F(x_c )}{F^{'''} (x_c ) x_c^3 },\theta= \frac{2F^{'}(x_c )}{F^{'''} (x_c ) x_c^2 }
\label{equi}
\end{equation}

\noindent The relations in Equation (\ref{equi}) are the same as those derived by Erez et al. \cite{erez2017criticality} using a somewhat different approach. At the critical point, $m=0$, i.e., $x_c=x^*$, the steady state at the SP bifurcation. Since $x^*$ is a triple root of $F(x^* )=0$, one gets $F(x_c )=0,F^{'} (x_c )=0,F^{''} (x_c )=0$. Thus from Equation (\ref{equi}), $\theta=0,h=0$ at the critical point. Analogously, from Equation (\ref{mag}), $X=0$ at the SP bifurcation point $R=0,H=0$. . Since in a MF theory the thermal fluctuations are ignored,  the MF Ising phase transition is similar to the SP bifurcation in a deterministic dynamical system. From Equation (\ref{ising}), the power-law singularities of various quantities, close to the critical point,  in terms of the critical exponents can be determined following standard procedure \cite{erez2017criticality,barenblatt1992n}. With $h=0$, the order parameter $m=0$ when the reduced temperature $\theta$ is $> 0$ and $m \sim \pm (-3\theta)^\beta$ for $\theta<0$. The value of the critical exponent $\beta=\frac{1}{2}$. For $\theta=0$, $m \sim (3h)^{1/\delta}$ with $\delta=3$. Similarly, the susceptibility $\chi \equiv (\partial_{h} m)_{h=0}$ exponents are $\gamma=1$ for both $\theta>0$ and $\theta<0$. Equivalent critical behaviour is exhibited by the dynamical quantities (Equation (\ref{imper})). The rate equations governing the dynamics of pitchfork bifurcation are system-specific but since the equations reduce to the same normal form in the vicinity of the bifurcation point, the critical behaviour is universal. For a one-variable dynamical system, one can define a bifurcation potential $U(x)$ as \cite{sornette2006transitions, strogatz2018nonlinear}
\begin{equation}
\frac{dx}{dt}=-\frac{dU(x)}{dx}
\end{equation}
The local minima of the potential correspond to the stable steady states. In the case of the SP bifurcation described by the normal form in Equation (\ref{sp_norm}), the bifurcation potential $U_{pitchfork} (x)=-\frac{R}{2}  x^2+\frac{x^4}{4}$   has a form identical to that of the Ginzburg-Landau potential \cite{barenblatt1992n} describing mean-field critical point transitions in statistical physics. In the region of monostability (bistability), the potential has one minimum (two minima). At the bifurcation point, the potential has a characteristic flat bottom. In the next Section, we extend the analogies between the MF Ising phase transition and bifurcation in a positive feedback model in terms of the phase diagram. The analogy is illustrated with a population biology model exhibiting the Allee dynamics.

\section{Phase diagram of single-variable dynamical model}
We consider a simple model of population dynamics exhibiting the Allee effect \cite{berec2007multiple, bose2017allee}. The differential rate equation in dimensionless form is given by
\begin{equation}
\frac{dx}{dt}=x(1-x)-\frac{bx}{a+x}
\label{allee}
\end{equation}

\begin{figure}
\begin{minipage}[c]{0.99\linewidth}
\begin{center}
\includegraphics[width=13.2cm]{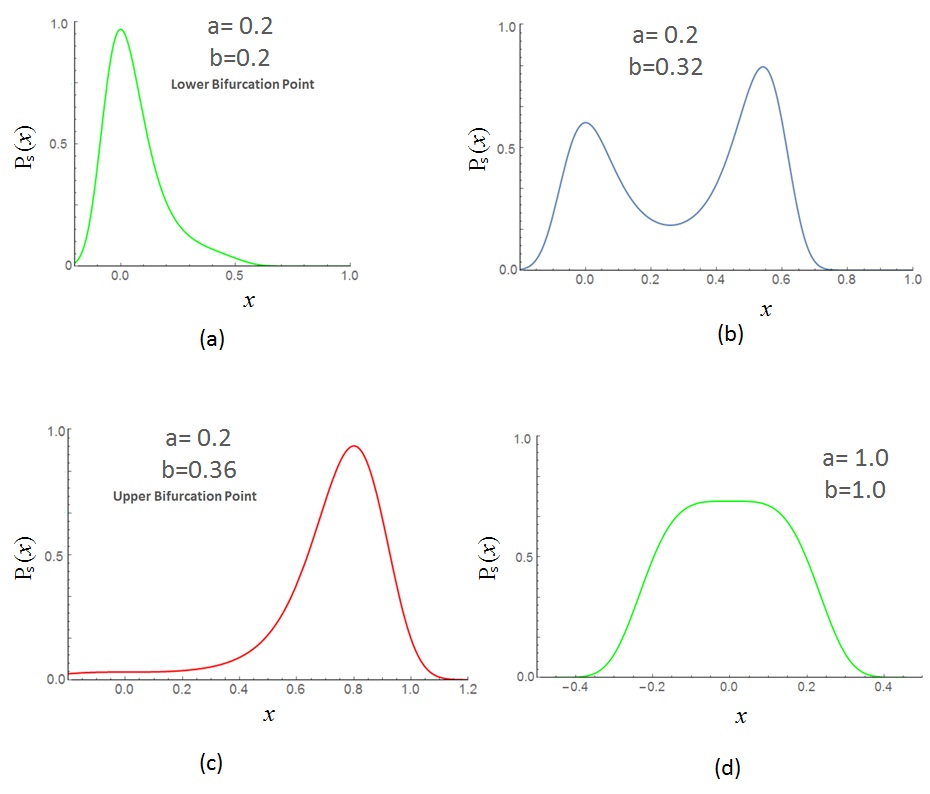}
\end{center}
\end{minipage}
\caption{Steady state probability distributions $P_s(x)$ versus $x$. (a)-(c) Steady state probability distributions corresponding respectively to the lower SN bifurcation point, the region of bistability and the upper bifurcation point. (d) Steady state probability distribution at the SP bifurcation point.}
\label{fp}
\end{figure}

\noindent The Allee effect denotes a negative per capita growth rate below a critical population density. In the course of time the population becomes extinct due to the negative growth rate. Since the model has a simple mathematical form, it is analytically tractable. The steady states are given by
\begin{equation}
x_1=0 ,x_2, x_3=\frac{1-a\mp\sqrt{(1+a)^2-4b}}{2}
\label{alsol}
\end{equation}
Figure 2 (left panel) shows the $b-a$ phase diagram with the shaded region indicating the region of bistability.  The equations defining the lower and upper boundaries of the bistable region are:
\begin{equation}
b=a ,b= \frac{(1+a)^2}{4}
\label{allee}
\end{equation}
respectively. The steady states on the lower boundary are $(0,0,1-a)$ and the same on the upper boundary are $(0,\frac{(1-a)}{2},\frac{(1-a)}{2})$. The double roots on the boundaries indicate SN bifurcations at which a stable steady state loses stability and there is a discontinuous jump from one branch of stable steady states to the other branch of such states. The discontinuous transition is analogous to a first-order thermodynamic phase transition in which a thermodynamic quantity, which is the first derivative of the free energy with respect to an appropriate variable, changes discontinuously.  The right panel of Figure 2 shows the plots of steady state $x_s$ versus the parameter $b$ for $a=0.2$ (upper plot) and $a=1.0$ (lower plot). In these plots, the solid (dashed) lines represent stable (unstable) steady states. The upper plot describes the SN bifurcation at which a discontinuous transition from a finite to zero population density takes place. The lower plot corresponds to SP bifurcation at $a=1,b=1$. There is a continuous transition to population extinction at this point. This is the well-known absorbing transition in which once the system reaches an absorbing state (zero population density) there is no way of coming out of the state. Outside the region of bistability in the $b-a$ phase diagram, one has regions of monostability: $x_1=0$ above the upper bifurcation line and $x_3$  (Equation (\ref{alsol})) below the lower bifurcation line. The coexistence of two stable  steady states in the region of bistability is reminiscent of the coexistence of two phases below the critical point in liquid-gas and paramagnet-ferromagnet phase transitions. The relevant quantities in Equation (10) can be computed, with $F(x)$ given by the r.h.s. of Equation (12), as $x_{c}=(ba)^{1/3}-a$,  $F' (x_{c} )=1+2a-3(ba)^{1/3}$  and $F''' (x_{c} )=-6/(ba)^{1/3}$. In the presence of noise (fluctuations), one has steady state probability distributions $P_s(x)$ as a function of $x$ instead of sharp stable steady states. The steady state probability distributions have been computed using the general stochastic formalism based on the Langevin and Fokker-Planck (FP) equations \cite{ghosh2012emergent}. A one-variable Langevin equation (LE) containing additive noise term is given by

\begin{equation}
\frac{dx}{dt}=f(x)+\epsilon(t)
\label{LE}
\end{equation}
where $\epsilon(t)$ represents Gaussian white noise with mean zero and correlations given by
\begin{equation}
\langle\epsilon(t)\epsilon(t')\rangle =  2D\delta(t-t')
\end{equation}
where $D$ is the strength of the additive noise. The first term $f(x)$ on the right hand side of the LE represents the deterministic dynamics with $f(x)=x(1-x)-\frac{bx}{a+x}$ (Equation (12)). The additive noise $\epsilon(t)$ represents noise arising from an external perturbing influence. The FP equation corresponding to the LE is
\begin{equation}
\frac{\partial P(x,t)}{\partial t}=-\frac{\partial}{\partial x}[f(x)P(x,t)]+D\frac{\partial^{2}}{\partial x^{2}}[P(x,t)]\end{equation}
The steady state probability distribution is given by
\begin{equation}
P_s(x)=\frac{N}{\sqrt{D}}\exp[\int^{x}\frac{f(y)dy}{D}]
\label{sspd}
\end{equation}
where $N$ is the normalisation constant. Equation (\ref{sspd}) can be recast in the form
\begin{equation}
P_s(x)=Ne^{-\phi_F(x)}
\end{equation}
with the stochastic potential
\begin{equation}
\phi_F(x)=\frac{1}{2}ln D-\int^{x}\frac{f(y)dy}{D}
\end{equation}
Figures 3(a)-3(c) show the steady state probability distributions $P_s(x)$ for $a=0.2$ and $b=0.2,0.32, 0.36$  corresponding respectively to the lower SN bifurcation point, the region of bistability and the upper SN bifurcation point. The value of $D$ is taken to be $0.001$. The steady state probability distribution $P_s(x)$ exhibits a characteristic flat top at the critical point $a=1,b=1$ (Figure 3(d)) \cite{erez2017criticality}.

\section{Two-variable dynamical model}

\begin{figure}
\begin{center}
\begin{minipage}[c]{0.99\linewidth}
\includegraphics[width=13.2cm]{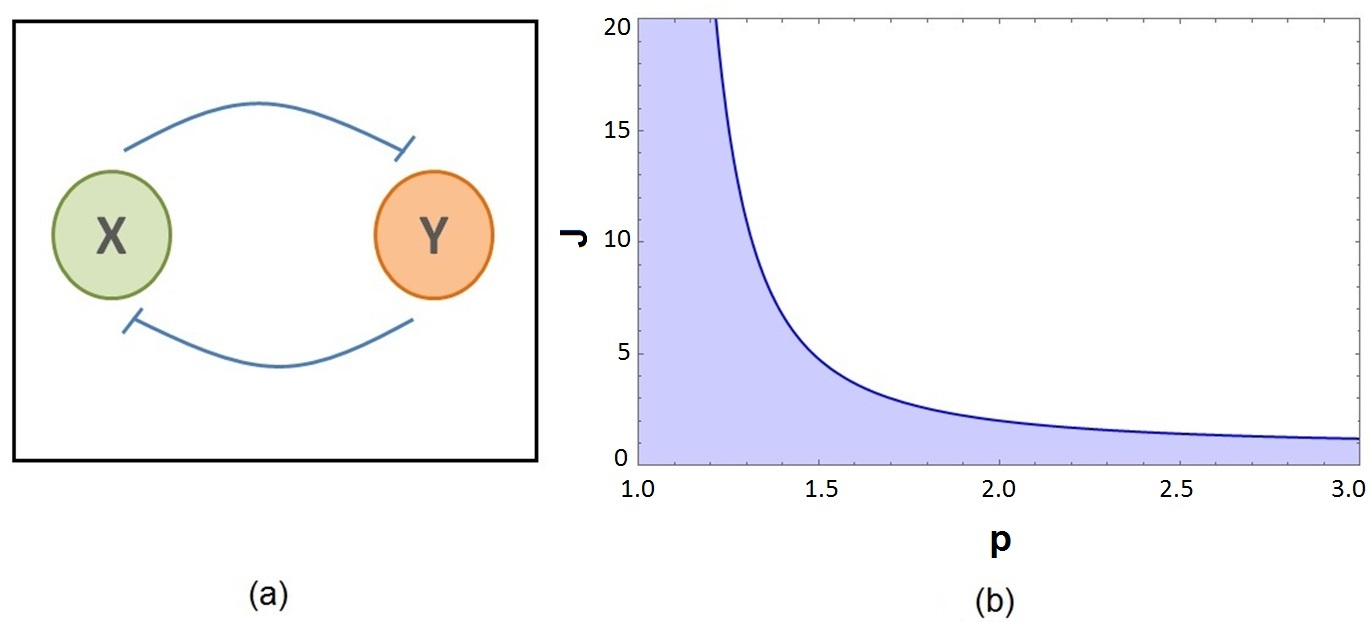}
\end{minipage}
\end{center}
\caption{(a) Genetic toggle. The protein products of genes $X$ and $Y$ repress each other's synthesis. The hammerhead symbol represents repression. (b) Plot of $J$ versus $p$ (Equation (26)). In the shaded region below the curve, there is only one equilibrium point whereas three equilibrium points exist in the parameter region above the curve. The SP bifurcation point is located on the curve itself.}
\label{togckt}
\end{figure}

In the case of one-variable dynamical systems, the normal form for SP bifurcation is well-known (Equation (\ref{sp_norm})). For more than one variable, there are very few studies on normal forms. Rajapakse and Smale \cite{rajapakse2016pitchfork} have proposed the normal forms  for SP bifurcation in the case of  a two-variable system which is, however, of limited applicability in the context of critical phenomena.  In this paper, we consider the two-variable genetic toggle \cite{gardner2000construction} and discuss the conditions under which a SP bifurcation is possible with equivalence between the bifurcation and MF Ising critical points. The genetic toggle, which functions as a switch,  consists of two genes the protein products of which repress each other's synthesis (Figure 4(a)). The feedback loop based on the mutual repression of two genes can be characterized as  a positive feedback loop and gives rise to two stable steady states OFF and ON in the appropriate parameter regime. In the OFF (ON) state, the concentration of a specific protein is low (high). The differential rate equations governing the dynamics of the toggle are, in dimensionless form, given by
\begin{equation}
\frac{dx}{dt}=\frac{J_x}{1+y^k}-x
\label{tog1}
\end{equation}
\begin{equation}
\frac{dy}{dt}=\frac{J_y}{1+x^s}-y
\label{tog2}
\end{equation}

\noindent where $x$ and $y$ are the protein concentrations, $J_x$ and $J_y$ the effective rates of protein synthesis and $k$, $s$ the cooperativity parameters of mutual repression. Pitchfork bifurcations occur at symmetric equilibria (steady states)  in systems satisfying specific symmetry conditions \cite{ellner2011dynamic}. The rate equations of the genetic toggle, for example, are symmetric under the interchange of $x$ and $y$ if the following conditions hold true: $J_x=J_y=J$ and $k=s=p$. A symmetric equilibrium solution $(x,x)$, is a solution of the single-variable  equation
\begin{equation}
f(x)=-x+\frac{J}{1+x^p}
\label{symmetric}
\end{equation}
Consider the nullcline curves
\begin{equation}
x=\frac{J}{1+y^p}=g(y),\: y=\frac{J}{1+x^p}=g(x)
\label{null}
\end{equation}
which are symmetric under the interchange of $x$ and $y$. The points of intersection of the curves are the steady states of the system. For arbitrary values of $J$ and $p$, one point of intersection always exists at the symmetric equilibrium $(x,x)$. In the case of a zero-eigenvalue bifurcation in a two-dimensional (2d) dynamical system, the nullclines always intersect tangentially, i.e., have the same slope at the bifurcation point. In the case of the SP bifurcation, the bifurcation point occurs at the symmetric equilibrium which is obtained as a solution of Equation (\ref{symmetric}). The condition for such a bifurcation is 
\begin{equation}
[g'(x) g'(y)]_{x=y}=1
\label{sp_cond}
\end{equation}
From Equations (\ref{symmetric})-(\ref{sp_cond}), one derives the following expressions for the symmetric equilibrium and  the SP bifurcation point:
\begin{equation}
x=y=(p-1)^{-\frac{1}{p}}  ,J=p(p-1)^{\frac{-(1+p)}{p}}
\label{rel}
\end{equation}
\begin{figure}
\begin{minipage}[c]{0.99\linewidth}
\begin{center}
\includegraphics[width=13.2cm]{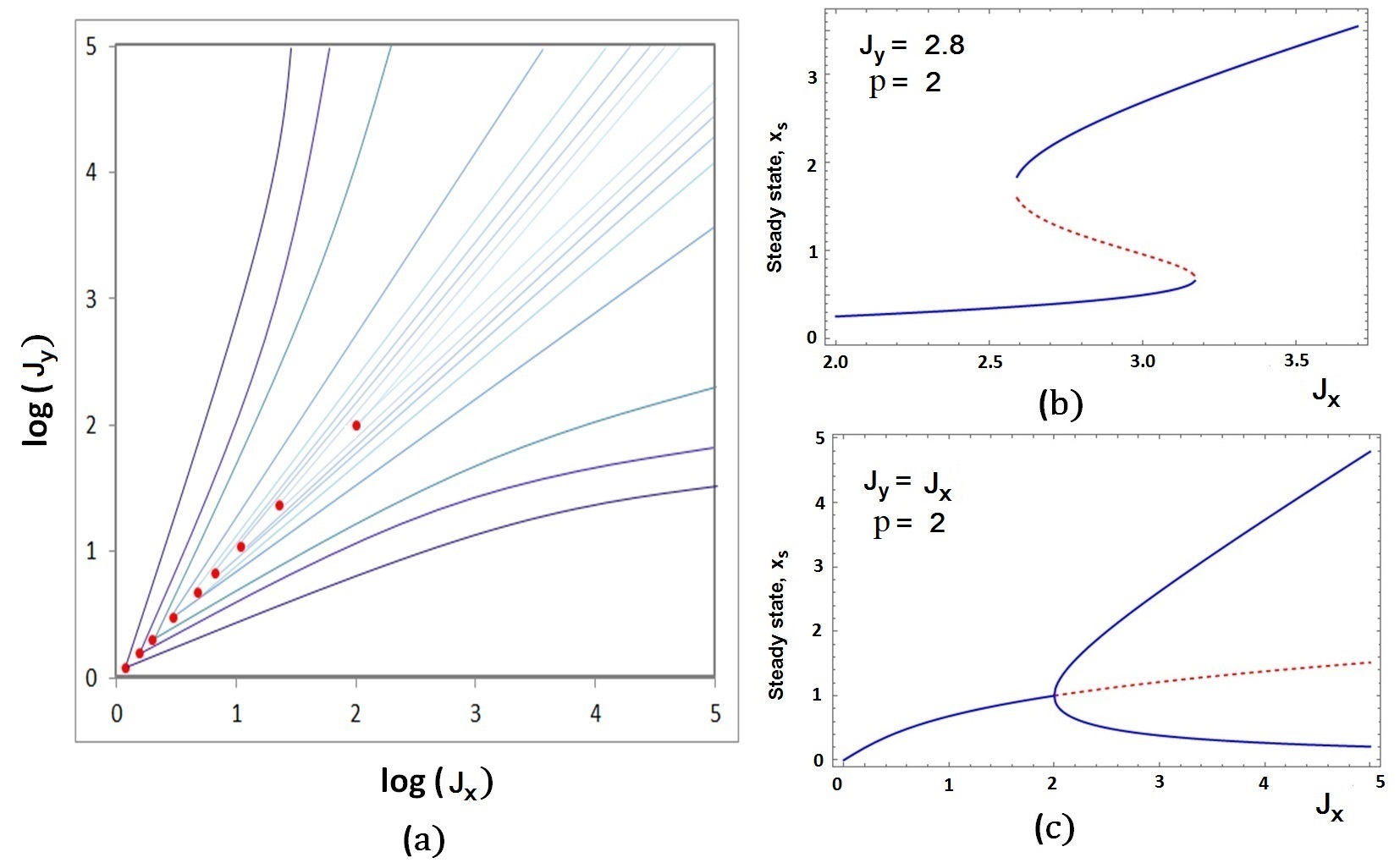}
\end{center}
\end{minipage}
\caption{(a) $\log J_y$ versus $\log J_x$ phase diagrams for $p=1.1,1.2,1.3,1.4,1.5,1.7,2.0,2.3$ and $3.0$ respectively. The regions of bistability are bounded by the solid lines which terminate at the SP bifurcation point calculated from Equation (\ref{rel}). (b), (c) Steady state plots of $x_S$ versus $J_x$ depicting the SN and SP bifurcations respectively. The solid (dashed) lines represent stable (unstable) steady states.}
\label{togphase}
\end{figure} 

\noindent The plot of $J$ versus $p$ (Figure 4(b)) defines the bifurcation curve below (above) which there is (are) one (three) equilibrium points. The SP bifurcation point is located on the curve itself. Figure  5(a) shows the $\log J_y$ versus $\log J_x$ phase diagrams for $p=1.1,1.2,1.3,1.4,1.5,1.7,2.0,2.3$ and $3.0$ respectively. The regions of bistability are bounded by the solid lines with higher values of $p$ yielding more extended regions of bistability. A region of bistability terminates at the SP bifurcation point (red points) at which the bifurcation parameter $J$ has the value as given in Equation (\ref{rel}). Figures 5(b) and 5(c) exhibit the steady state plots of $x_S$ versus $J_x$ depicting the SN and SP bifurcations respectively. The former occurs at points on the boundaries of the bistable region and the latter occurs at the SP bifurcation point, the termination point of the boundary lines. \\
To explore the critical behaviour of the two-variable dynamical model, the steady state equation is expressed in terms of a single variable. In the case of the symmetric toggle, the steady state conditions are derived from Equations (\ref{tog1}) and (\ref{tog2}) with  $J_x=J_y=J$ and $k=s=p$. As an illustration, consider $p=2$ keeping the parameter $J$ free. The single-variable steady state equation is obtained by substituting the steady state value $y=\frac{J}{1+x^2}$ in the steady state equation $x=\frac{J}{1+y^2}$. One thus obtains the single-variable steady state equation 
\begin{equation}
F(x,J)=x^5-J x^4+2x^3-2 J x^2+(1+J^2 )x-J=0
\label{poly}
\end{equation}
Following the same procedure as in Section 2, one finds the value of $x_c$ from a solution of  $F'' (x_c,J)=0$. The relation between the Ising and toggle parameters are as in Equation (\ref{equi}). From Equation (\ref{rel}), the value of the SP bifurcation parameter $J =2$  for $p=2$ and the equilibrium is $x=y=1$. One can check that all the three quantities,  $F(x_c ),F' (x_c ),F'' (x_c )$, are zero in this case. The main conclusion is that the symmetric two-variable genetic toggle and the MF Ising model belong to the same universality class in terms of the critical point singularities. To generalise, for any multivariable dynamical system exhibiting a SP bifurcation, the trick is to reduce the steady state equations to a single equation involving only one variable in order to demonstrate that the model belongs to the MF Ising universality class. As pointed out before, symmetric dynamical systems with the potential for bistability (for which positive feedback is a requirement) are candidate systems exhibiting SP bifurcation.

\section{Concluding Remarks}

The similarity between bifurcations and phase transitions has been pointed out in a number of studies on dynamical systems \cite{erez2017criticality, sornette2006transitions, qian2016framework,munoz2018colloquium, lesne2011scale,sole2011phase}. The SN and SP bifurcations are analogous to first and second-order (critical point) phase transitions in equilibrium systems. In a dynamical system, the asymptotic limit of time $t \longrightarrow \infty$  is equivalent to the thermodynamic limit required for the emergence of critical point singularities in equilibrium phase transitions.  A few recent studies have drawn analogies between the SP bifurcation point and the MF Ising critical  point \cite{erez2017criticality, sornette2006transitions, qian2016framework}. The significance of the study carried out by Erez et al. \cite{erez2017criticality} rests on the fact that it extends the MF Ising universality class of equilibrium thermodynamic models to include biochemical models with positive feedback describing nonequilibrium phenomena. In Section 2 of our paper, we have rederived some of the earlier results using a simple deterministic approach. This forms the basis of an easy-to-implement general formalism for exploring critical phenomena in biochemical systems with positive feedback. The method is illustrated in Section 3 with a model of population dynamics exhibiting the Allee effect for which the exact bifurcation diagram has been computed. The bifurcation diagram, analogous to the phase diagram of equilibrium phase transitions, identifies the different dynamical regimes of monostability and bistability. The boundaries of the region of bistability represents lines of SN bifurcations which terminate at the SP bifurcation point. The phase diagrams of equilibrium phase transitions like liquid-gas and paramagnet-ferromagnet share  a similar feature in terms of  a line of first order transitions terminating at a critical point. We have introduced stochasticity (additive noise) in the dynamics of the Allee model and shown in Figure 3  the evolution of the steady state probability distribution as a bifurcation parameter is varied to bring about SN bifurcations. The flat-topped distribution in Figure 4 provides the signature of a  SP bifurcation point which is a critical point.  The Allee model explains qualitatively the experimental observations of Dai et al. \cite{dai2012generic} on laboratory populations of budding yeast \cite{bose2017allee}. The Allee effect was realized experimentally through the cooperative growth of yeast in sucrose. The experimentally obtained bifurcation diagram for the population density was of the SN type with a discontinuous transition at the bifurcation point from a finite population density to population extinction. Appropriate redesigning of the experimental protocol could pave the way for observing critical phenomena in a laboratory population. \\
\noindent In Section 4 of our paper, we have proposed an approach for studying Ising-like critical phenomena in multivariable dynamical systems exhibiting SP bifurcations. The strategy in this case is to reduce the number of steady state equations to a single one involving only one variable. As an illustration of the method we considered the two-variable genetic toggle, one of the most well-known of synthetic biology circuits. Assuming a symmetric genetic toggle, the reduction to a one-variable steady state equation is possible. Again, the theoretically predicted critical behaviour can be tested in actual experiments. In a biological system, the responses to stimuli can be either graded or binary \cite{bose2012culture}. In graded response, the response changes continuously as a parameter indicative of the stimulus amount is changed till a saturation level in response is attained. In binary response, the response in an individual cell is either low or high with very few cells associated with intermediate level response. An increase in the stimulus amount has the sole effect of raising the fraction of the cell population with high level of response. In  a phase diagram (Figures  2(a) and 5(a)), a graded response is obtained if the path of stimulus changes bypasses the region of bistability. Traversal through the wedge-shaped region brings about a binary response. In the lactose utilization network, the graded as well as  binary responses have been observed through the appropriate tuning of biochemical parameters \cite{ozbudak2004multistability}.\\
\noindent Rapid advances in experimental techniques in the last decade or so have made it possible to probe a variety of novel biological phenomena involving positive biochemical feedback \cite{pomerening2008uncovering,bose2012culture,mitrophanov2008positive,pal2014non}. Multistability is a possible outcome of dynamics controlled by one or more positive feedback loops.  Bifurcations resulting in bistability constitute a universal theme in cell biological processes. Erez et al \cite{erez2017criticality} derived relations between the  parameters of a biochemical system with positive feedback and thermodynamic quantities like  effective temperature, magnetic field and order parameter. It is thus possible to extract the effective thermodynamic quantities from biological data. This was demonstrated by Erez et al. using fluorescence data from mouse T cells without the requirement of fitting and in the absence of  knowledge of the underlying molecular details. The interpretation of the data in terms of the effective parameters shows that positive (negative) values of the effective temperature θ bring about monostability (bistability). Positive (negative) values of the effective magnetic field give rise to high (low) steady state levels. There is significant  experimental scope for the interpretation of the effective thermodynamic quantities including susceptibility and specific heat in bifurcation-related biological phenomena. The divergence of ``susceptibility" could provide definitive evidence of  a SP bifurcation point in an experiment. In recent studies \cite{pal2014non,bose2017criticality,mojtahedi2016cell}, cell differentiation has been proposed as a critical point transition. The idea has considerable experimental support \cite{bose2017criticality,mojtahedi2016cell} and theoretical studies based on a positive feedback model provide evidence \cite{pal2014non,bose2017criticality} of a flat topped steady state probability distribution close to the pitchfork bifurcation point. A large number of studies have been carried out in recent years on the early signatures of regimes shifts at bifurcation points \cite{scheffer2009early,pal2013early}. These signatures, which include critical slowing down, rising variance and increasing  lag-1 autocorrelation function, are generic to any type of bifurcation point and the bifurcation points have often been termed as ``critical points" in existing literature. The analogy with the Ising critical point singles out the pitchfork bifurcation point as the true ``critical point" characterised by power-law singularities and a steady state probability distribution with a flat top. Biological systems offer an unprecedented opportunity to test the concepts of dynamical systems theory and statistical physics in actual experiments. The expansion of the mean-field Ising universality class from equilibrium models of phase transition to nonequilibrium models of biological systems highlights the universal features of such systems close to criticality. The study of critical phenomena in living systems is a newly emerging area of research which is expected to witness a surge in activity in the coming years \cite{munoz2018colloquium}.  

\section*{Acknowledgement}
\noindent IB acknowledges the support  by CSIR, India, vide sanction Lett. No. 21 (0956)/13-EMR-II  dated 28. 04. 2014.
\section*{References}

\bibliographystyle{ws-ijmpc}
\bibliography{ws-sample.bib}

\end{document}